# A CONCEPTUAL METADATA FRAMEWORK FOR SPATIAL DATA WAREHOUSE

M.Laxmaiah[1] and A.Govardhan[2]

[1]Department of Computer Science and Engineering,
Tirumala Engineering College, Bogaram (V),
Keesara (M), Hyderabad, AP, India-501301
laxmanmettu.cse@gmail.com
[2]Department of Computer Science and Engineering,
Jawaharlal Nehru Technological University,
Hyderabad, AP, India-500085
govardhan_cse@yahoo.co.in

***ABSTRACT:***

*Metadata represents the information about data to be stored in Data Warehouses. It is a mandatory element of Data Warehouse to build an efficient Data Warehouse. Metadata helps in data integration, lineage, data quality and populating transformed data into data warehouse. Spatial data warehouses are based on spatial data mostly collected from Geographical Information Systems (GIS) and the transactional systems that are specific to an application or enterprise. Metadata design and deployment is the most critical phase in building of data warehouse where it is mandatory to bring the spatial information and data modeling together. In this paper, we present a holistic metadata framework that drives metadata creation for spatial data warehouse. Theoretically, the proposed metadata framework improves the efficiency of accessing of data in response to frequent queries on SDWs. In other words, the proposed framework decreases the response time of the query and accurate information is fetched from Data Warehouse including the spatial information.*

***KEYWORDS:***

*Data Warehouse, Metadata, Geographic Information Systems, Spatial Data Warehouse, Global Positioning Systems, Online Transaction Processing, Online Analytical Processing.*

## 1. INTRODUCTION

The Metadata is a most basic prerequisite key component of the data warehouse. The Metadata is the central point of tracking, designing, building, retrieving information in data warehouses. It defines and describes both the business and information view of the information repository and provides the information from Data Warehouse. It also assists end users in locating, understanding and accessing information in applications. The challenging task for data warehouse users and professionals is to keep track of the *data* and about *data warehouse keys*, *attributes* and *layout changes* for the longer period of time [9].

Data warehousing applications are based on high performance databases that allows client /server architecture to integrate diverse data types in real world [27].Over 80 percent of business data have some spatial context such as Zip code, customer address, or store location [6]. By using technology that integrates this spatial component with data warehouse, organizations can unveil the hidden potential in their data. Proposed design and implementation of data marts results in more organized data structure, better integration of disparate data, spatially enabled data analysis, reduced decision making time, and improved decisions. These operational data includes manufacturing, payroll, inventory, and accounting that are designed to allow users to run a business, but do not analyze it. Operational systems data tend to be process oriented and narrowly focused. For example in a retail business operational data include items such as point of sale and inventory control captured by bar code which includes location information either by street address, Zip code or Area code. These data must be integrated with standard attribute data into an enterprise wide repository from which decision makers can perform ad-hoc analysis and run reports. The new dimension i.e location information gives decision makers more definition of their data and allows them to ask new questions about relationships in their database.

The classes of queries that are needed to support the decision making are difficult on spatial data bases which include geographical information. This initiated the attention of the assessors and developer towards the spatial data mining. Spatial data warehouses are combination of traditional data warehouses with spatial information resulting Spatial Data Warehouse (SDW). Spatial data warehouses are based on the concept of the data warehouses and additionally support to store, index, and aggregation and analyze spatial data. The operations of standard data warehouses are also extended to spatial data warehouses. The characteristics of a spatial data warehouse include *conceptual models*: star and snow flake schema for spatial attributes. *Spatial components*: spatial measures, spatial dimensions, and spatial hierarchies[26], spatial OLAP operations, efficient query processing and optimized memory usage.Thefigure1 shows the taxonomy of data warehouse, spatial data warehouse and spatiotemporal data warehouse.

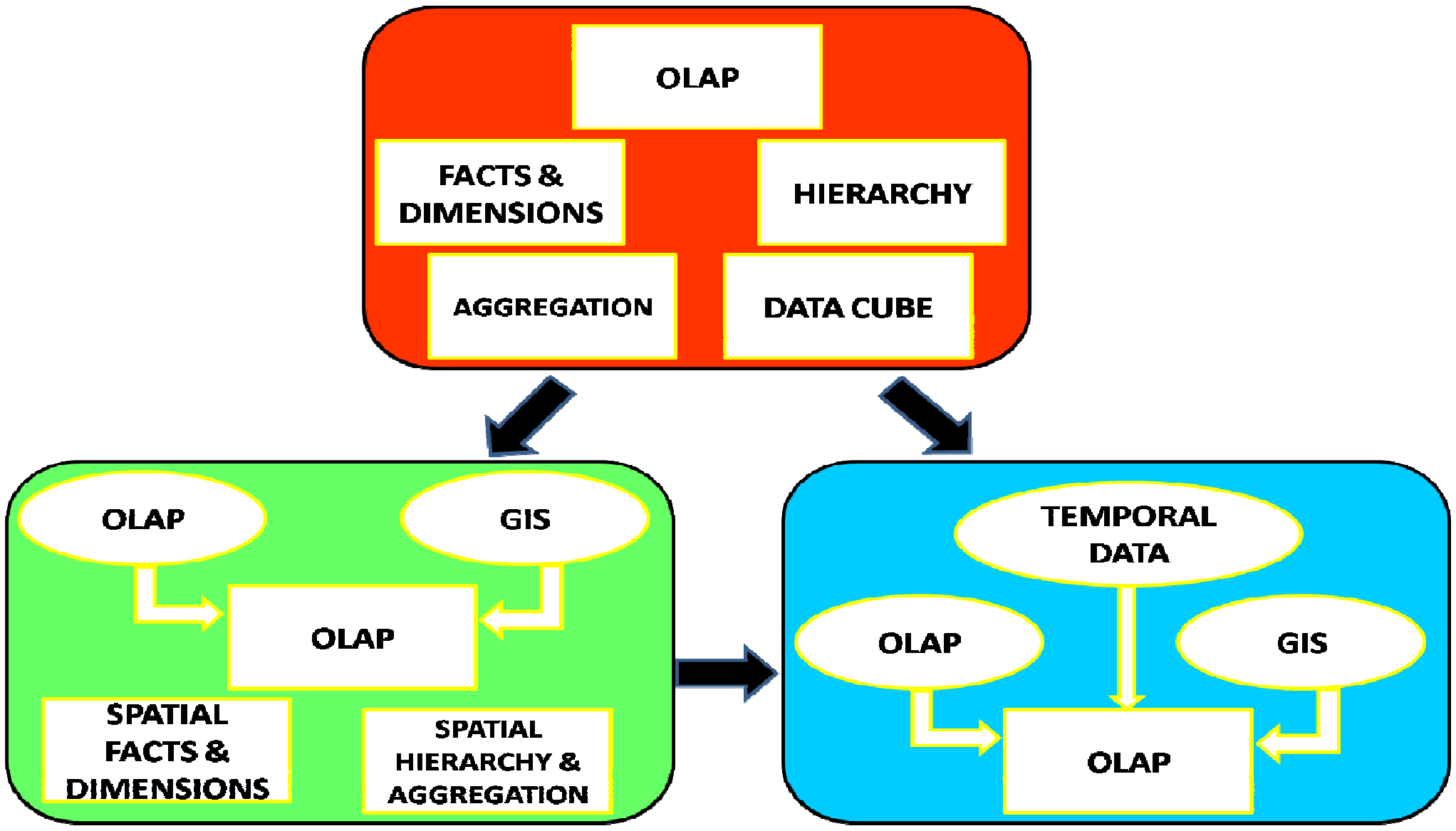

Figure 1: Taxonomy of spatial and spatiotemporal data warehouse

## 1.1 Spatial Data

Huge amount of GIS data is available on paper in printed form but not in digital form and is to be transformed into the digital representations. This captured data is used for multiple applications like Archeological Analysis [ 28], Marketing Research [29 ]; Urban Planning [ 30]. Data already produced in digital formats will certainly ease the work and speeds up the process of developing the GIS applications where huge amount of the data is accessed and analyzed.

*Global Positioning System (GPS)* information can also be collected and then imported into a GIS. A current trend in data collection gives users the ability to utilize in field computers with the ability to edit live data using wireless connections. This has been enhanced by the availability of low-cost mapping-grade GPS units with decimeter accuracy in real time. This eliminated the need to post process, import, and update the data in the office after fieldwork has been collected. This improves the ability to incorporate positions collected using larger range finders. New technologies also allow users to create maps as well as analysis directly in the field, making more efficient projects and more accurate mapping [11].

*Satellite Remote Sensing* provides another important source of spatial data. Here satellites use different sensor packages to passively measure the reflectance from parts of the electromagnetic spectrum or radio waves that were sent out from an active sensor such as radar. Remote sensing collects *raster data* that can be further processed using different bands to identify objects and classes of interest, such as land cover [16].

## 1.2 Schema and Data Integration

Integration of the data is often different between applications and databases in the way data is stored, Semantics of the data and the way the data is organized. Two types of heterogeneities are there in the data integration: *Schema Heterogeneity and Data Heterogeneity*.

***Schema Heterogeneity:*** Multiple sources of data must be combined to retrieve information that is not contained entirely in either one. Spatial data is captured from various sources and stored in various places, so they do not have the same schemas, to overcome this problem they specify a Schema Model or Standardized Data for Integration [4].

***Data Heterogeneity:*** Data integration is done either by the user or by integrating the databases directly. The user integration is done by specifying some constraints or physically sorting the data and integrating it, where as in integrating databases technique the data is moved into one database. Since databases are normally developed independently, integrating different database schema is becoming complex because of different structures, terminologies and focuses used by different data base designers. In case of GIS, managing heterogeneity among the data is very complex because it must take into account non spatial and spatial data[21]. Most common conflicts encountered at this stage are different data formats for the same field in semantic, missing values, abbreviation data, and duplicated data to face the problems of the database and GIS interpretability and methodologies defined on them. Conflicts at statistical and semantic level have to be dealt carefully for better data integration.

## 2. RELATED WORK

Data warehousing has captured great attention of practitioners and researchers for a long time ago, whereas the aspects of spatial data collection, integration and metadata is one of the crucial issues in spatial data warehousing [5].The formats of the spatial data and non spatial data are different and leads to the problem of data integration. Constructing an integrated spatial data warehouse from existing databases and data warehouses will create the problem of interoperability among the data warehouses and its users. This process prevents the scalability of the resultant schema and data warehouses [1,2,3].To work with this data, the users need appropriate descriptions and additional knowledge about the data. This data is naturally called as Metadata [7,10,12,13].

Distributed geographic information services are one of the possible solutions for the management of very large scale GIS databases. However, it is currently difficult to access distributed GIS datasets and web mapping services [24] remotely due to their heterogeneity. Many research projects, including digital libraries, data clearinghouses and data mediators are focusing on the management issues of distributed geographic information services. Currently, one of the popular solutions is to create metadata associated with geospatial data items and services, which can be interpreted by users or metadata search engines.

Metadata becomes the key to bridge the heterogeneous environments of distributed GIS databases and services and to provide users with the semantics and Syntactic of GIS databases [14]. However, by collaborating with operational metadata contents and hierarchical metadata repositories, the new metadata framework can help users and systems to access on-line geo-data objects, software components, and web map services efficiently.

Metadata directs SDW to extract heterogeneous data from different sources, GIS databases and also directs to the application system to obtain the digital geographical product and provides users some services about the geographical product. The Figure 2 depicts the Metadata of Spatial Data Warehouse and Relative Technologies.

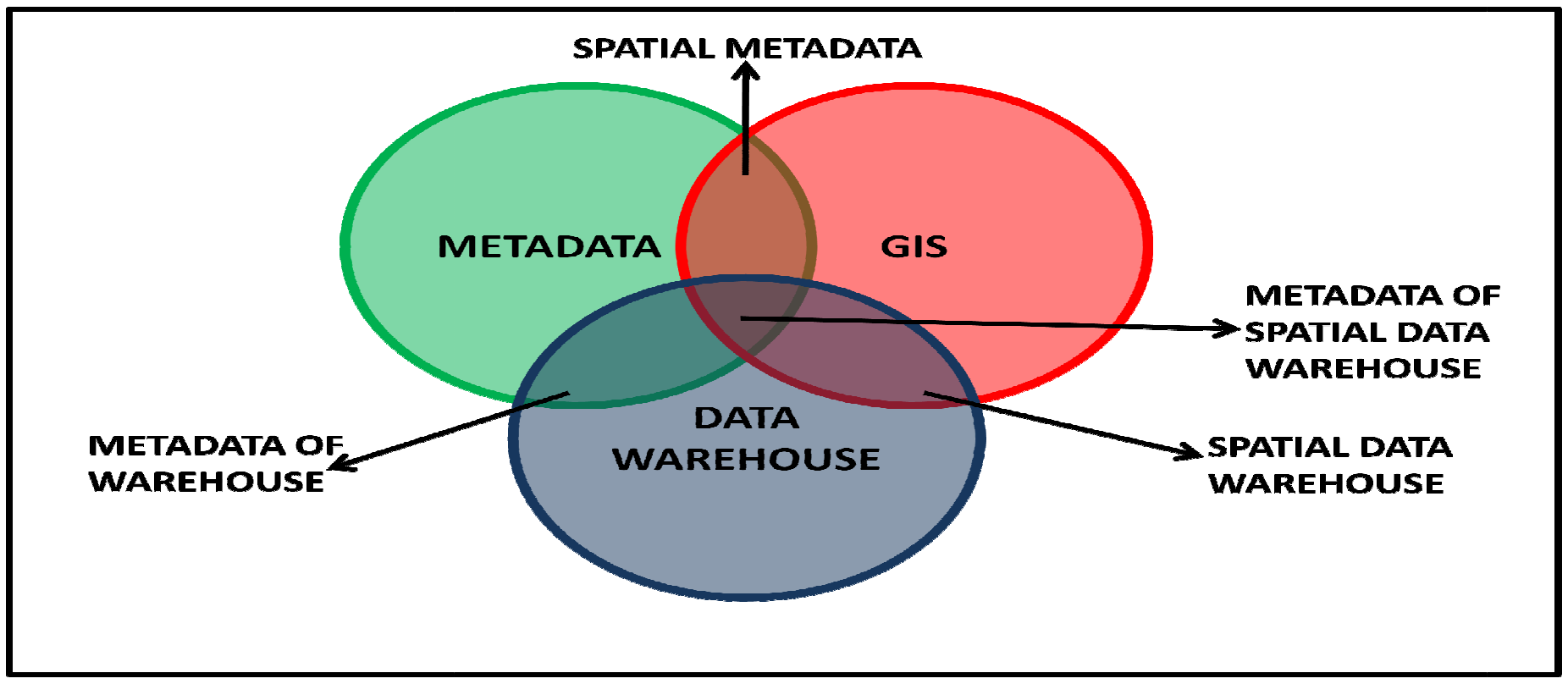

Figure 2: Metadata of Spatial Data Warehouse and Relative Technologies

## 2.1 Data warehouse

Data warehouse is an effective database management and application system that stores; there is a large amount of complex data which comes from different sources, in different formats, types and models stored into data warehouse [16]. In general, the main task of database is to provide services for online transaction processing (OLTP). It's consistency, standard; transaction processing and data capacity to be saved is relatively small. The data warehouse stores the historical, integrated and unaffected data sets for Online Analytical Processing (OLAP).The data set is often modeled multi-dimensional representation to meet data cube requirements.

## 2.2. Metadata

Metadata is data about data; book tag, version, and edition publication are the metadata for the data book in the library. It provides easy inter communication between data producer and user in readable and understandable format. The difficulty arises when metadata is transferred into digital data, it is not easy to manage and apply metadata. The challenges faced are expressed as follows:

- No in place expedient tool to select data sets from various databases.
- No unique technique is available for data sets in order to access these data sets.
- Not knowing how to understand and transform data sets when user wants to apply these data sets.
- No known affiliation information on data production, update and distribution.

Metadata can be used to many fields such as enterprises of data document, data distribution [31], browsing data and data transformation [14] and so on. There is a firm relationship between metadata and data content to be illustrated by metadata.

## 2.3. Geographic Metadata

When geographic information emerges in digital form, various new challenges come into existence for management and application of geographic data [22]. These challenges are included as follows but not limited to;

- In order to manage and maintain mobility of geographic data it is necessary for data producer to process a tool.
- To have the lowest effect for production and maintenance of geographic data, if data producer and indicator exchanges to built data document in order to save some technology information with geographic data.
- User need to know the efficiency approach, how to select geographic data? and need to know the place how to find geographic data for user application?.

When users want to make some digital product with geographic data, they need to understand geographic data and transform geographic data into own approach oriented format. It is very important to have geographic metadata for relating content, quality, status etc. Geographic Metadata can provide characteristic information of spatial data sets for the generalization and abstraction of spatial data character. The difference between geographic metadata and metadata is that there is a lot of information of spatial location in geographic metadata.

## 3. CONCEPTUAL METADATA FRAMEWORK FOR SPATIAL DATA WAREHOUSE

The design of conceptual metadata for GIS data objects needs to consider what kind of operations are associated with geo-data objects. The most important four representative tasks are proposed for specification of the spatial data warehouse. Metadata may consist of *Map display*, *spatial query*, *Spatial Operations*, and *Data connectivity* [15]. The real development of object metadata can include more tasks or elements based on specific task requirements or application needs. The conceptual metadata model of SDW is depicted in figure 3 below;

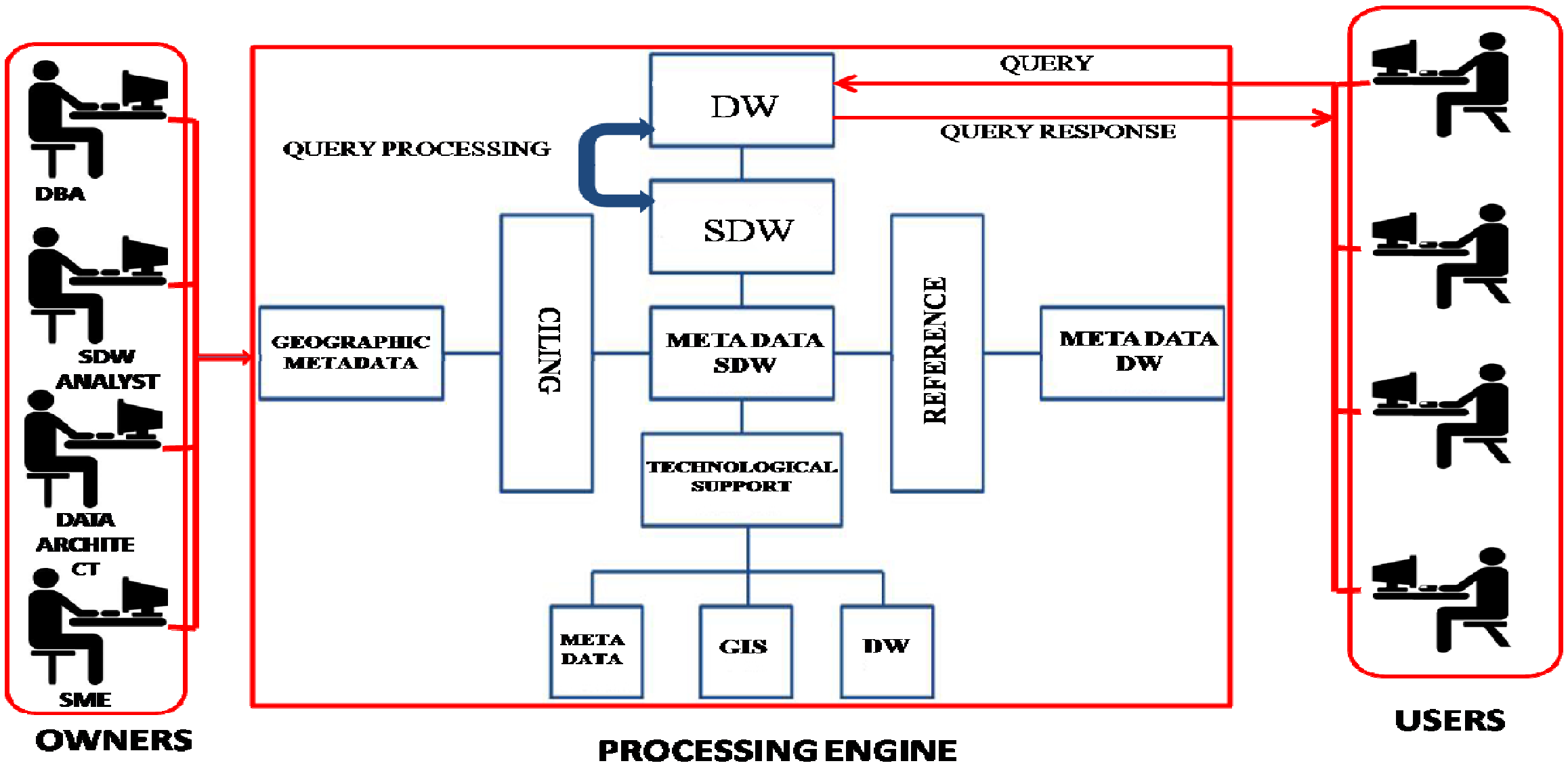

Figure 3: Conceptual Framework Metadata for Spatial Data Warehouse

The design of the *map display* metadata element is to specify the representation methods of geo-data objects on electronic media or computer screens. The contents of *map display* metadata includes: *type of features* (raster/vector, point, line, polygon, or volume); *attribute type* (nominal, ordinal, interval or ratio, or multiple attributes); *map symbols used* (attribute lookup table, symbol size, symbol icons/shapes); *color schemes* are required (2 bits-B/W, 8bits-256 color, 32 bits-true color),and scale threshold. These metadata contents can be interpreted automatically by mapping software to apply both a color schemes and map symbols dynamically during the given time [8, 14].

With the help of *map display* metadata, geo-data objects become self describable and self manageable map layers. Web map users can decide whether they want to change color schemes manually and symbols or just adopt the default settings configured in the map display metadata .One actual objective to keep in the mind at this stage is that the definition of map symbol should consider the dynamic environment of distributed mapping services with different computer display techniques and screen resolutions. For example, if a line symbol is displayed on tablet PC with small screen resolution (300x200), the width of the line symbol will be adjusted automatically according to the size of screen.

The design of spatial query metadata is to describe the GIS query requirements of geo-data objects[18]. It includes what kind of query language used (natural language, SQL or other spatial query languages)?; What is the query syntax?, the query interface required (the interface which can provide remote access point) and results display spatial operation on metadata will specify the possible spatial operations associated with geo-data objects and their requirements.

Data connectivity metadata focus on the mechanisms of remote access and download procedures for geo-data objects. The design of data connectivity metadata will specify the interactions between Geo-data objects and remote machines or databases [19].The data connectivity takes care of what type local access methods used (the communication in a single machine)?; What are the remote access methods (remote database connections)?;What type data compression/un-compression methods used *(wavelets, gzip compression*)?; and registration of data objects.SDW should be efficient enough in updating the spatial query of users from various sources and workstations.

## 3.1 Geographic Information System

A geographic information system (GIS) is a system designed to capture, store, manipulate, analyze, manage, and present all types of *geographical data [21].*GIS is a growing multidisciplinary technology based on geography, computer science, and sociology and so on. Various application domains quietly used ranging from economical, ecological and demographic analysis, city and route planning.GIS is basically composed of four major parts:

***Spatial data capturing:*** Entering information into the GIS[25] consumes much of the time of it's practitioners. There are a variety of methods used to enter data into a GIS where it is stored in a digital format. Existing data printed on paper or Polyethylene terephthalate (PET film) maps can be digitized or scanned to produce digital data. A digitizer produces vector data as an operator traces points, lines, and polygon boundaries from a map. Scanning a map results in raster data that could be further processed to produce vector data. Survey data can be directly entered into a GIS from digital data collection systems on survey instruments using a technique called coordinate geometry. Positions from a global navigation satellite system like Global Positioning System can also be collected and then imported into a GIS.

***Representation of spatial data:*** GIS data represents real objects such as roads, land use, elevation, trees, waterways etc, with digital data determining the mix. Real objects can be divided into two abstractions: *discrete objects* (e.g;a house) and *continuous fields* (such as rainfall amount, or elevations). Majorly, there are two broad methods used to store data in a GIS for both kinds of abstractions mapping references: *raster images* and *vector*. Points, lines, and polygons are the stuff of mapped location attribute references. A new hybrid method of storing data is that of identifying point clouds, which combine three-dimensional points with RGB information at each point, returning a "3D color image". GIS thematic maps then are becoming more and more realistically visually descriptive of what they set out to show or determine.

***Visualization of spatial data**: Cartography* is a visual representation of spatial data[23]. The vast majority of modern cartography is done with the help of computers, usually using GIS but production of quality cartography is also achieved by importing layers into a design program to refine it. Most GIS software gives the user substantial control over the appearance of the data. Cartographic work serves two major functions: First, it produces graphics on the screen or on paper that convey the results of analysis to the people who make decisions about resources. Wall maps and other graphics can be generated, allowing the viewer to visualize and thereby understand the results of analyses or simulations of potential events. Web Map Servers facilitate distribution of generated maps through web browsers using various implementations of web-based application programming interfaces (AJAX, Java, Flash, etc.).

***Analysis of spatial data***: GIS spatial analysis[32] is a rapidly changing field, and GIS packages are increasingly including analytical tools as standard built-in facilities, as optional toolsets, as add-ins or 'analysts'. In many instances these are provided by the original software suppliers (commercial vendors or collaborative non commercial development teams), whilst in other cases facilities have been developed and are provided by third parties.

## 3.2 Metadata of Data Warehouse

The lifecycle of metadata is divided into three phases*: collection, maintenance* and *equipmen*t. These three phases encourage each other in order to play metadata an important role in data warehouse [17].

The ***collection phase's*** main job is to identify metadata and input the metadata into central repository. The collection of metadata should be done automatically as much as possible so that there is higher reality for collection of metadata. The collection of metadata can be done automatic, but some metadata has to be collected manually.

During ***maintenance phase*** metadata must keep track of actual change of data. For example if the structure of relational database table changes the metadata should describe the table change and must be updated in order to image change and it must be done to have metadata in good condition in DW and make the data available for analysis purpose. It will bring very high level data automatically for physical metadata imaging the structure of data source and data warehouse.

***Equipment phase*** is to provide proper metadata and tools which can be applied on. It is the phase to yield after paying out a lot of work in the phases of *collectio*n and *maintenance* metadata. The important key factor is to equip metadata to match correct metadata and specifically according to users demand.

## 3.3 Spatial Data Warehouse

When the GIS information is maintained in DW it is called as SDW. The sharing and operations on each spatial information, the analysis and generalization of spatial information become very important in general research of geographic problem. In general, GIS is spatial oriented-application and is group by work flow. The data in GIS is often in original state. The function of GIS is only process operations of adding, deleting, and modifying to data and simple spatial selection and spatial analysis. Keeping the interest of the business in the mind a unique format is saved in SDW. Stability development in an interesting and challenging procedure to deal with multi-store and multi formats of data in SDW and it is transformed to main source.

## 4. CONCLUSION

The data warehouse ensures efficient management and operations on geographic data at a cost of consuming least time in replying spatial queries, difficulty in optimizing and storing the spatial information in distributed sources and irrelevant representations of spatial information available in SDW.It will be better to employ SDW with a require way of storing the information despite optimize various sources in various formats. The most common drawback of the SDW usually is inefficient replies to the user queries becoming the distributed metadata of DW, SDW. Uniqueness in SDW would be speed up the replying to the user queries. In this paper, we present the concept of introducing the spatial data warehouse and spatial metadata concepts for better query processing.

The conceptual metadata framework for SDW presented in this paper provides a well development and explains the accessing aspects of the information in SDW. It will also arrange a better result for long waited query replies and it will improves the efficiency of the SDW updates and replies intern helps the business to understand the needs and design new policies and schemes. In this paper we present the need of the unique type of SDW which is also a distributed in nature but *store the information in its unique way* which intern helps the system to process queries efficiently well and make use of best technical resources. The data captured from various sources and bring various equipments to store various formats and type of data which could lead to accessing of the data makes time consuming process while replying queries. Providing a dynamic SDW which captures the information from different sources in different formats but stores in its own uniquely designed standards will reply the time consuming queries. The proper capturing, maintaining and equipment is improved a lot and show it's awesome.

## AUTHORS

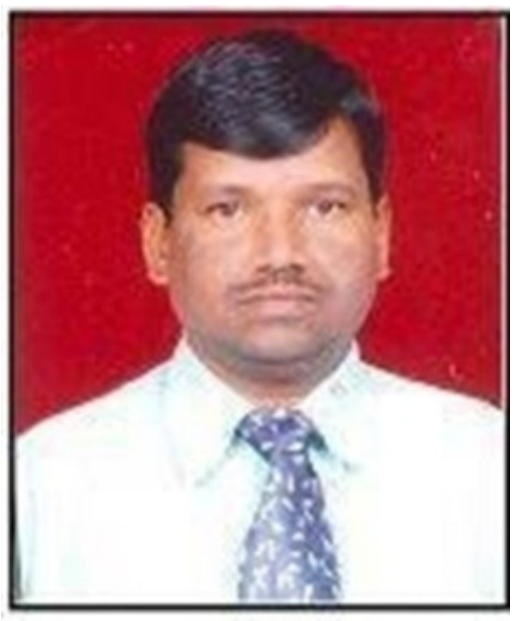

**Mr.M.Laxmaiah** is research scholar in Jawaharlal Nehru Technological University, Kukatpally Hyderabad. He is currently working as Professor and Head of CSE Department in Tirumala Engineering College, Bogaram (V), Keesara (M), Hyderabad, AP, India. He has 15 Years of experience in Education and 4 Years of experience in Research field. He has 5 research Publications at International Journals. His areas of interest include Database Management Systems, Compiler Design and Data warehousing & Data Mining.

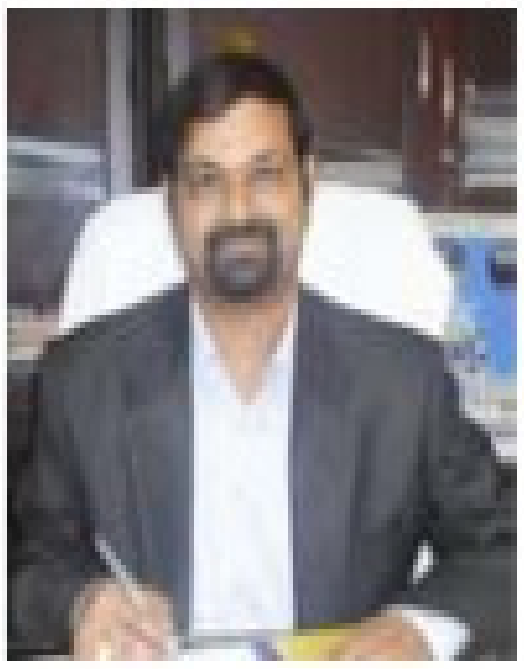

**Dr.A.Govardhan** did his BE in Computer Science and Engineering from Osmania University College of Engineering, Hyderabad in 1992.M.Tech from Jawaharlal Nehru University, Delhi in 1994 and PhD from Jawaharlal Nehru Technological University, Hyderabad in 2003.He is presently Director of Evaluation, JNTUH, Kukatpally, Hyderabad. He has guided more than 120 M.Tech projects and number of MCA and B.Tech projects. He has 180 research publications at International/National Journals and Conferences. His areas of areas of interest include Databases, Data Warehousing &Mining, Information Retrieval, Computer Networks, Image processing and Object Oriented Technologies